\pdfoutput=1
\documentclass[11pt,a4paper]{article}
\usepackage{graphicx}

\newcommand{\Sol}{$_\odot$}
\newcommand{\arcsec}{{\hbox{$^{\prime\prime}$}}}

\newcommand{\aap}{A\&A}
\newcommand{\apj}{ApJ}
\newcommand{\jgr}{J. Geophys. Res.}
\newcommand{\pasj}{PASJ}
\newcommand{\apjl}{ApJL}

\begin{document}

\begin{center}
\Large
Microwave and EUV Observations of an Erupting Filament and Associated 
Flare and CME

\bigskip
\large
C. E. \sc{Alissandrakis}\footnote{
Department of Physics, University of Ioannina, GR-45110 Ioannina, Greece},
A. A. \sc{Kochanov}\footnote{
Institute of Solar-Terrestrial Physics, Lermontov St. 126, Irkutsk 664033,
Russia},
S. \textsc{Patsourakos}$^1$, A. T. {\sc Altyntsev}$^2$,\\ S. V. {\sc Lesovoi}$^2$, 
N. N. {\sc Lesovoya}$^2$\\

\end{center}

\smallskip
\centerline{Accepted for publiction in PASJ, 2013/07/17}


\normalsize
\section*{Abstract}
A filament eruption was observed with the Siberian Solar Radio Telescope (SSRT) 
on June 23 2012, starting around 06:40 UT, beyond the West limb. The filament could be 
followed in SSRT images to heights above 1 R\Sol, and coincided with the core of 
the CME, seen in LASCO C2 images. We discuss briefly the dynamics of the eruption:
the top of the filament showed a smooth acceleration up to an apparent velocity of 
$\sim1100$\,km\,s$^{-1}$. Images behind the limb from STEREO-A show a two 
ribbon flare and the interaction of the main filament, located along the primary 
neutral line, with an arch-like structure, oriented in the perpendicular 
direction. The interaction was accompanied by strong emission and twisting motions. 
The microwave images show a low temperature component, a high temperature 
component associated with the interaction of the two filaments and another high 
temperature component apparently associated with the top of flare loops. We 
computed the differential emission measure from the high temperature AIA bands 
and from this the expected microwave brightness temperature; for the emission 
associated with the top of flare loops the computed brightness was 35\% lower 
than the observed. 

\section{Introduction}
Magnetic structures containing dense and cool 
chromospheric plasma which is suspended against the solar gravity high up in the
corona are observed as filaments against the solar disk or as prominences 
off-limb. They can experience long periods of stability by persisting
for several solar rotations. 

\begin{figure*}
\begin{center}
\includegraphics[width=16cm]{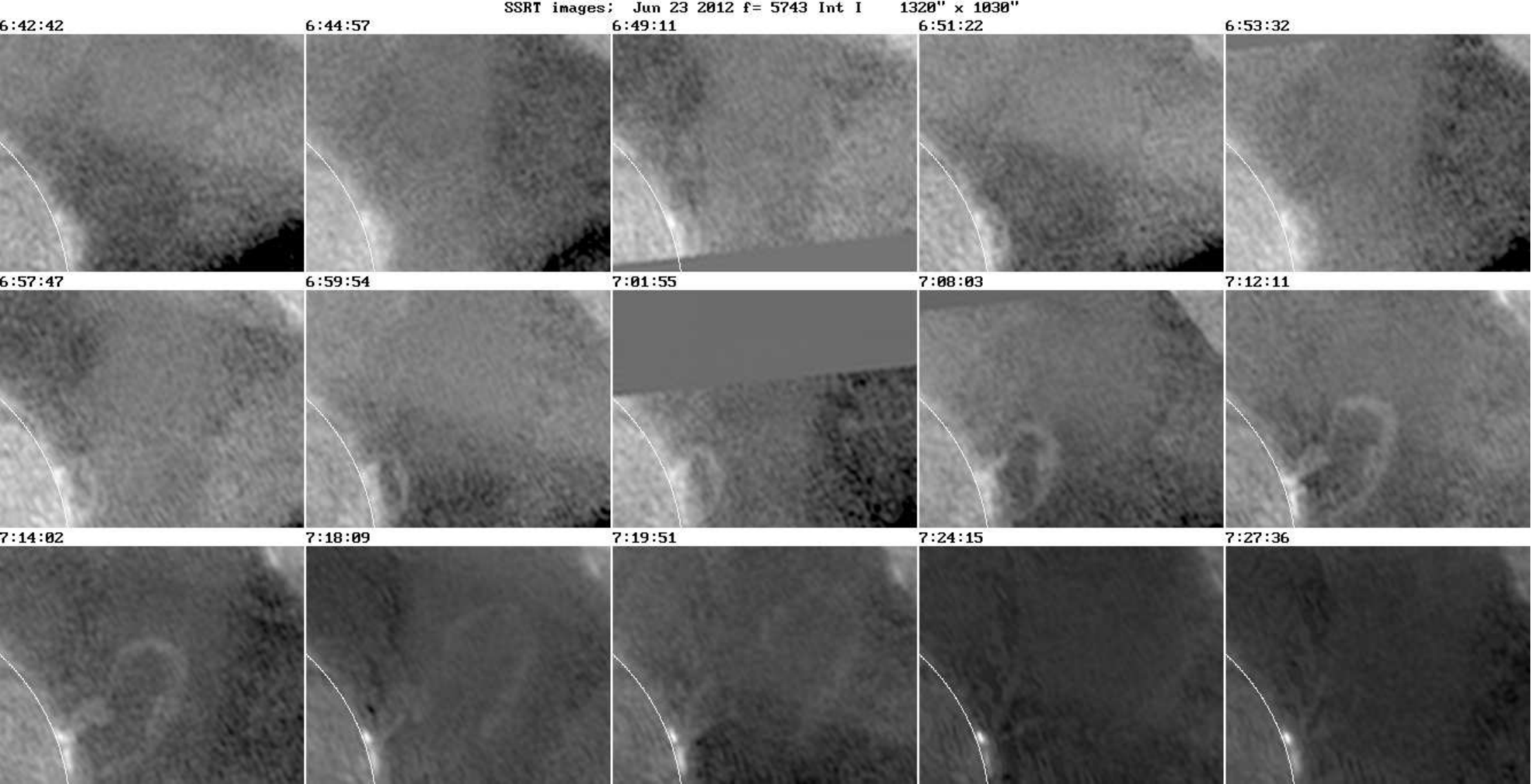}
\end{center}
\caption{Sequence of SSRT images during the filament eruption. The white arc shows 
the position of the photospheric limb. Note that the SSRT does not observe the 
entire sun simultaneously, thus the times marked correspond to the top of the 
filament. In this and subsequent images, solar north is up, solar west to the 
right.}
\label{SSRT}
\end{figure*}

The occasional eruption of quiescent filaments has been known for many decades 
(see {\it e.g.\/} D'Azambuja (1925)). Soon after Coronal Mass Ejections 
(CMEs) were detected, it was realized that filament eruptions can lead to CMEs 
with a high probability; for example, Gopalswamy et al. (2003) found that 72\% 
of the filament eruptions they studied led to CMEs. Moreover, filament eruptions 
are involved in a significant fraction of the observed CMEs. For example 
Subramanian \& Dere (2001) showed that 60\% of CMEs in their sample was associated 
with filament eruptions. Sometimes, during their ascend, eruptive filaments 
experience some mild heating ({\it i.e.\/}, not leading to flare temperatures), 
{\it e.g.\/}, Landi et al., 2010). CMEs related to filament eruptions 
could be launched either from active regions (active region filaments) or the 
quiet Sun (polar crown filaments).

An important problem of the physics of CMEs is related to the nature of their 
seed magnetic structure with basically two opposing schools of thought. There are 
CME models which require the existence of a magnetic flux rope, i.e., a structure 
with twisted magnetic field lines along its central axis, before the eruption, 
whereas there are other models in which a flux rope forms during the eruption 
(see for example the reviews of Forbes, 2000 and 
Klimchuk, 2001). While direct imaging observations of pre-existing flux 
ropes started to become available with recent AIA observations in hot, flare 
emissions ({\it e.g.\/}, Zhang et al. ,2012, Patsourakos et al., 2013),
observations of filaments both before and during CMEs can also supply important 
information for the above-mentioned problem. This is because filaments frequently 
show evidence of twisted structure before and during their eruption and undergo 
kinking and rotations ({\it e.g.\/} Williams et al., 2005). All the above are 
frequently interpreted in terms of instabilities of magnetic 
flux ropes. It is thus obvious why the study of filaments, both when they are
in quiescence and when they erupt, is important in the study of solar eruptions.

In the microwave range filaments have been observed both on the disk as 
temperature depressions and as prominences in emission beyond the limb 
(Kundu, 1972). Combined microwave (VLA) and EUV (SUMER, CDS) 
observations of a filament on the disk were analyzed by 
Ciuderi Drago et al. (2001). In the metric range filaments are seen as 
depressions on the disc ({\it e.g.\/} Marqu\' e, 2004), but there have 
also been reports of their association with emission sources at decametric 
wavelengths (Lantos et al., 1987).

With the availability of daily observations from the Nobeyama Radio Heliograph 
(NoRH) and the Siberian Solar Radio Telescope (SSRT), several reports of erupting 
filaments and associated CMEs have been published (see, {\it e.g.\/} 
Gopalswamy, 1999). Uralov et al. (2002) 
presented an event observed by both instruments and proposed that the eruption 
may have been caused by the interaction of two filaments. More cases were studied 
by Grechnev et al. (2006).

In this work we report a filament eruption observed with the SSRT. Unfortunately 
the event did not occur within the time range of the NoRH, so we have no spectral 
information. However, we complete our microwave 
data with images from STEREO behind the limb and high cadence AIA/SDO images; both 
STEREO and SDO data are used in this context for the first time. We start with an 
overview of the event, we discuss briefly its dynamics and we use AIA data to 
compute the contribution of the high temperature plasma to the microwave emission

\section{Overview of the Event}\label{over}
The Siberian Solar Radio Telescope (SSRT, Grechnev et al., 2003), operating 
at 5743 MHz (5.2\,cm), observed a filament eruption beyond the West limb, which 
started around 06:40 UT on June 23, 2012. Figure \ref{SSRT} gives a sequence of 
SSRT images, which show the classic picture of initial slow rise followed by a 
rapid expansion. It is the central part of the filament that rises, while its 
edges remain more or less fixed. Although its contrast decreases with time, the 
filament remains visible up to a height of more than 1 R\Sol, until its top 
overlaps with a grating image of the solar disk, seen in Figure \ref{SSRT} at the 
upper right corner of the images. The visibility of the filament is also affected 
by sidelobes from the two bright points that appeared near the limb during the 
late phases of the event; most of this effect has been removed by ``cleaning'' in 
Figure \ref{SSRT}. The brightness temperature of the filament was $\sim13000$\,K 
at 06:53:22 UT, dropping to $\sim3500$\,K at 07:19:51 UT; note that due to the 
varying sky background the accuracy of these measurements is not better than 
500\,K. 

\begin{figure*}
\begin{center}
\includegraphics[width=\textwidth]{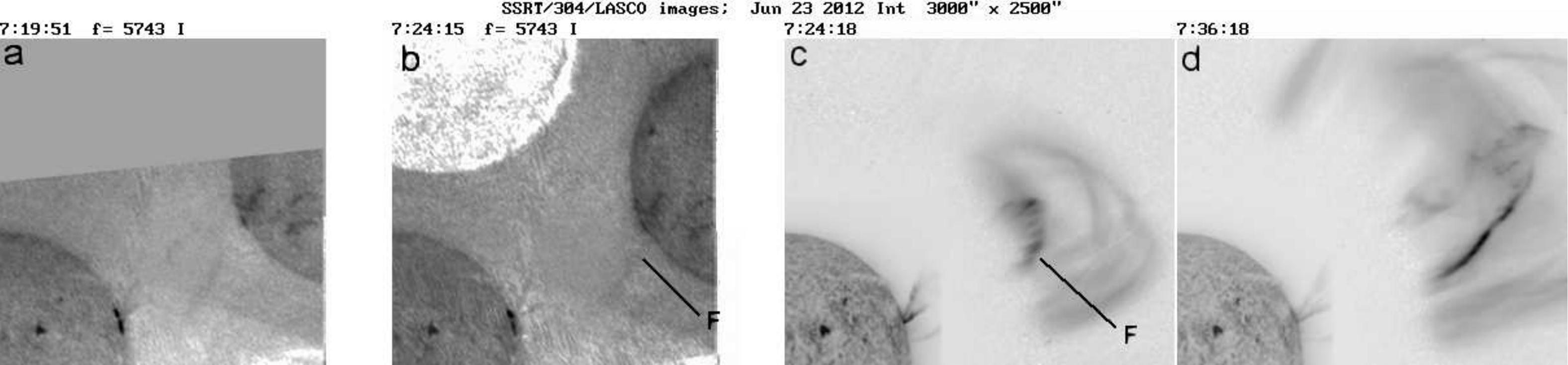}
\end{center}
\caption{SSRT (a, b) and composite LASCO/C2 - AIA 304 \AA\ (c, d) images. The intensity 
before the event has been subtracted in the LASCO images. F points to the flank 
of the filament, seen both in the SSRT and the C2 image at 07:24 UT.}
\label{C2}
\end{figure*}

Figure \ref{C2}c shows the first LASCO C2 image (at 07:24:18 UT) where the CME 
went above the occulting disk, together with the corresponding 304 \AA\ AIA image. 
In the nearest SSRT image (Figure \ref{C2}b) at 07:24:15 UT, the filament is very 
faint and its top is already inside the grating image of the sun; still, the 
flank of the filament (marked F in the figure) is visible in both the SSRT and C2 
images, which demonstrates that the erupting filament corresponds to the core of 
the CME. For reference, the previous SSRT and the next C2 images are shown in 
Figure \ref{C2}a and \ref{C2}d respectively. 

\begin{figure*}
\begin{center}
\includegraphics[width=\textwidth]{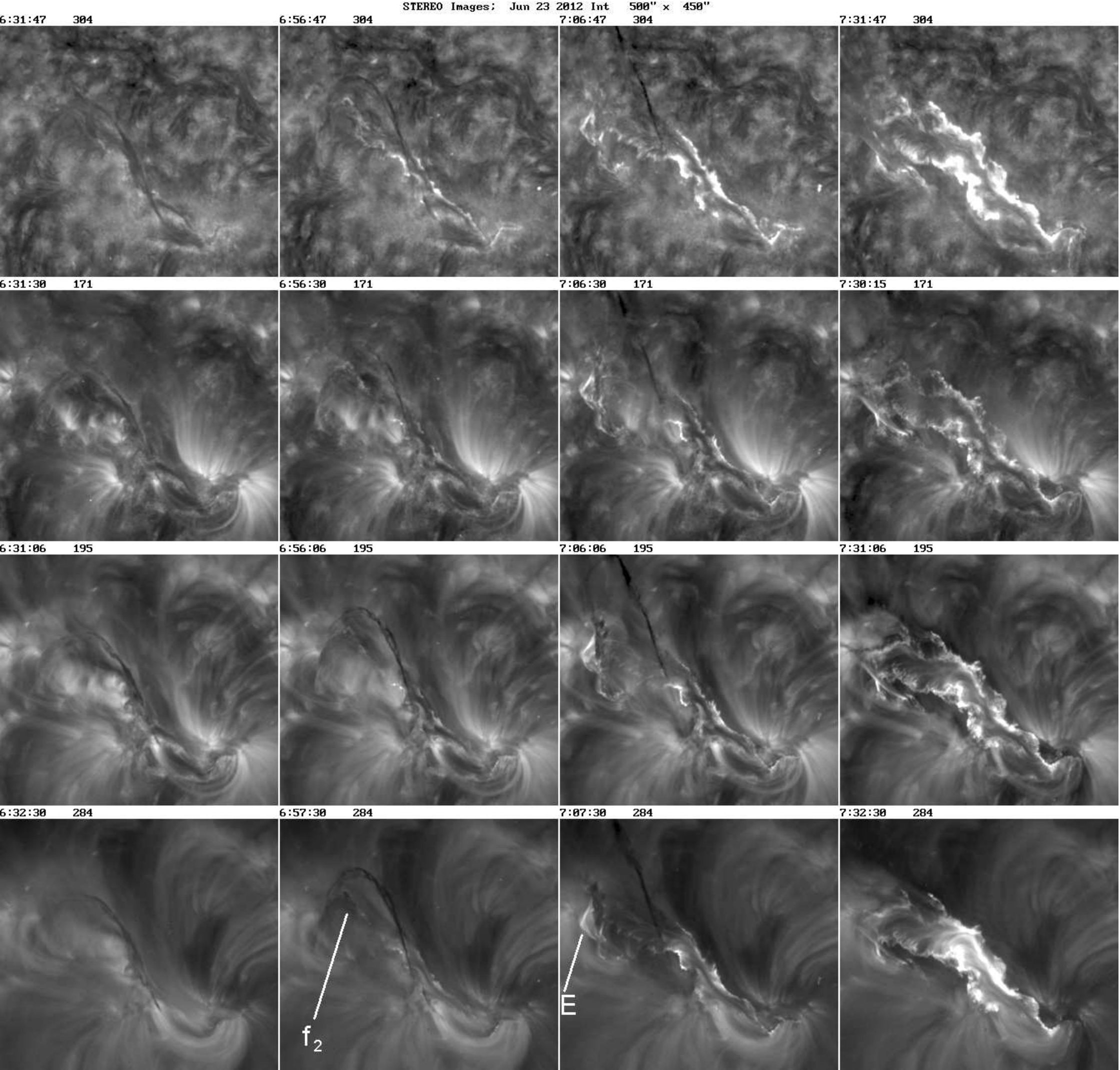}
\end{center}
\caption{STEREO-A images of the event in the 304, 171, 195 and 204 \AA\ bands. 
f$_2$ marks the second filament and E the bright emission which appeared as a result 
of its interaction with the main filament. For the full sequence of images see 
movie 3}\label{stereo}
\end{figure*}

It is clear from Figure \ref{SSRT} that the event occurred behind the limb. In 
the days prior to the eruption, the filament was visible on the disk, on top of
the neutral line of NOAA Active Region 11506; this filament had merged with a 
pre-existing filament located at its NE. AR 11506 was in its decay phase, with no 
sunspots visible after June 18.
 
At the time of the event STEREO-A was at an angle of 116 degrees with respect to 
the earth. As expected, STEREO-A/EUVI gives a very good view of what happened behind the 
limb (Figure \ref{stereo}; see also movies 1 and 2 which show almost simultaneous 
SDO and STEREO-A images in the 304 and 171 \AA\ bands, as well as movie 3 showing 
the flare evolution in all four STEREO bands). We can see here a classic 
two ribbon flare, accompanying the filament eruption, with the first brightening 
appearing around 06:47 UT at 304\,\AA\ and at 06:50 at 171\,\AA. Note that it was the eastern 
portion of the filament that first erupted. Note also the existence of a second, 
arch-like dark filament, best seen in the 195 and 204 \AA\ bands (marked f$_2$ in the Figure), oriented almost 
perpendicularly to the main filament. While the eruption of the main filament was 
already in progress, it interacted with the main filament, giving rise to intense 
emission in all STEREO bands with a maximum around 07:07 UT (third column of 
images in Figure \ref{stereo}, marked E in the 284 \AA\ image). We further note 
that the eruption had the characteristics of a {\it disparition brusque}: 
the main filament reformed approximately one day later, as evidenced by the 
STEREO-A data.

\begin{figure}
\begin{center}
\includegraphics[width=.5\textwidth]{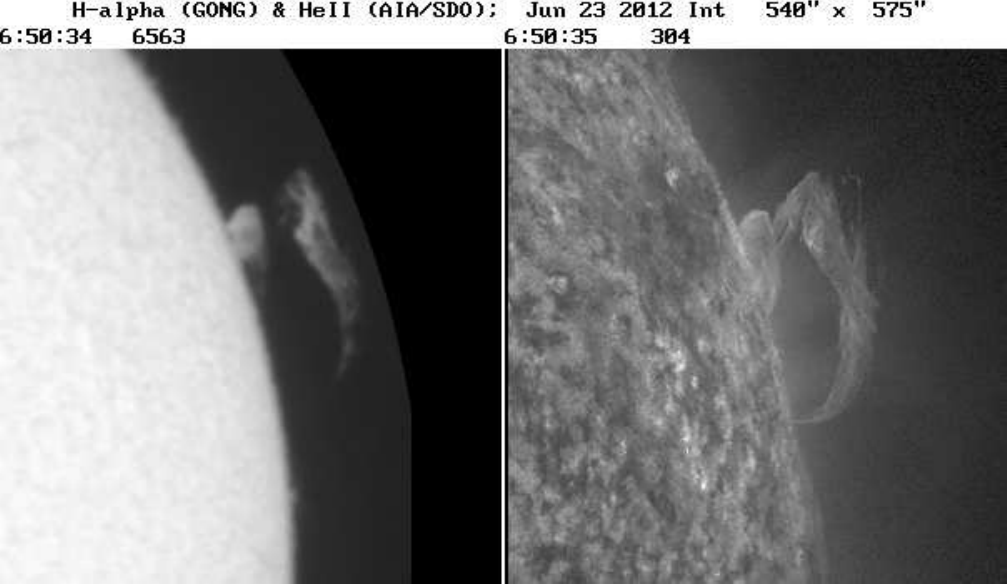}
\end{center}
\caption{H$\alpha$ image from the GONG network (left) and He{\sc ii} 304 \AA\ image
from AIA/SDO (right) during the early stage of the eruption.}
\label{GONG}
\end{figure}

Going back to the earth point of view, Figure \ref{GONG} shows an H$\alpha$ and a 
He{\sc ii} 304 \AA\ image in the early stage of the event, both at a resolution 
much better than the 23\arcsec\ resolution of the SSRT. The images, H$\alpha$ in 
particular, show two discrete prominence structures: a lower structure, apparently 
corresponding to filament f$_2$ seen in STEREO-A and a higher structure, 
corresponding to the main erupting filament (we will call this structure f$_1$ 
hereon). A detailed examination of the series of images shows that the activation 
of f$_2$ occurred after f$_1$ (see movie 1 and discussion in section \ref{dynamics}).
Also visible in the SDO images in the 171\,\AA\ band (movie 2) is a set of 
expanding loops spanning the filament.

\begin{figure}[h]
\begin{center}
\includegraphics[width=.8\textwidth]{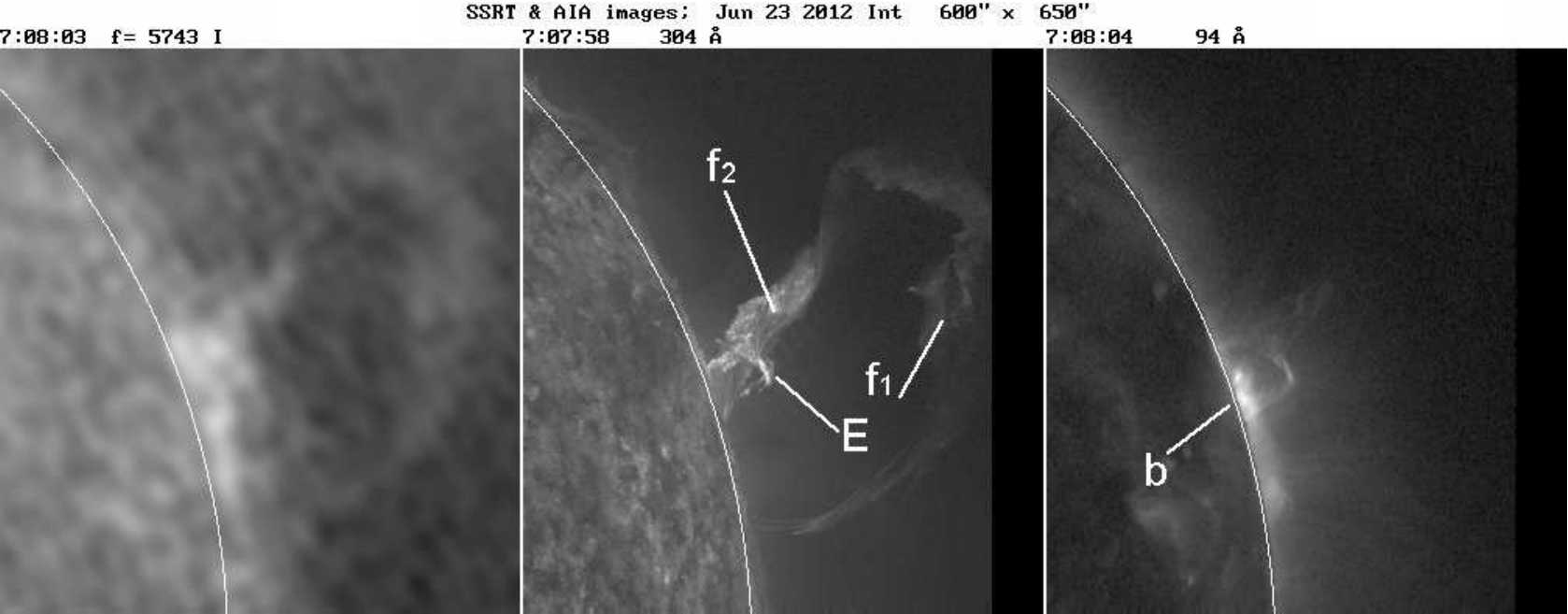}
\end{center}
\caption{Comparison of an SSRT image (left) with AIA/SDO images in the 304 and 94 
\AA\ bands.}
\label{AIA}
\end{figure}

Let us now compare the morphology of the microwave and the EUV emission. We remind 
the reader that the AIA channels are sensitive to temperatures in the range from 
$\sim10^5$\,K to $\sim10^7$\,K, while the microwave brightness temperature is 
much lower. Still a comparison is meaningful, because in the microwaves we may 
have contribution of optically thin emission from material at temperatures much 
higher than the brightness temperature. We chose to compare in Figure \ref{AIA} 
an SSRT image with a low temperature band (304 \AA, $T\sim10^5$\,K) and a high 
temperature band (94 \AA, $\sim10^6$ and $\sim10^7$\,K) image. The time of the images in the figure was chosen so that the erupting 
filament was still inside the AIA field of view, which does not extend in height 
beyond 0.35 R\Sol, significantly less than the SSRT.

Taking into account the difference in resolution, the SSRT image looks very much 
like the 304 \AA\ image; we note that the 304 \AA\ image is also very similar to the 
1600 \AA\ band image (not shown here), which has significant contribution from the 
C{\sc iv} lines at 1548 \AA, also formed at a temperature of $\sim10^5$\,K. 
Comparing the 304 and 94 \AA\ images, we can identify hot and cold components: The 
filaments f$_1$ and f$_2$ are practically invisible in the 94 \AA\ image, while the 
bright region E, the result of the interaction between the two, is seen in both 
bands. In addition, the high temperature band shows a bright component just above 
the limb (marked b in the figure), invisible in the 304 \AA\ image. 

It is interesting to note that all these components are present in the SSRT 
image. On the basis of the above, we can roughly identify three temperature 
components in the microwave emission: (a) a low temperature component, associated 
with the erupting filaments, (b) an intermediate temperature component associated 
with the interaction of the two filaments and, (c) a hot component just above the 
limb, the nature of which will be further elaborated upon later in this article. 
As for the filament-associated component, the fact that it appears in absorption 
on the disk implies that its electron temperature is below the disk brightness 
temperature of $\sim2\times10^4$\,K, {\em i.e.\/} much below the average formation temperature 
of the He{\sc ii} 304 \AA\ line. However, it is not clear if the appearance of the 
filament in the 304 \AA\ band image is due to the extension of its temperature 
sensitivity down to $10^4$\,K or due to the transition region between the filament 
and the corona.

\section{Radio Emission at Metric and Longer Wavelengths}
Earth-based instruments show no trace of metric emission associated with the event,
\begin{figure}[h]
\begin{center}
\includegraphics[width=.8\textwidth]{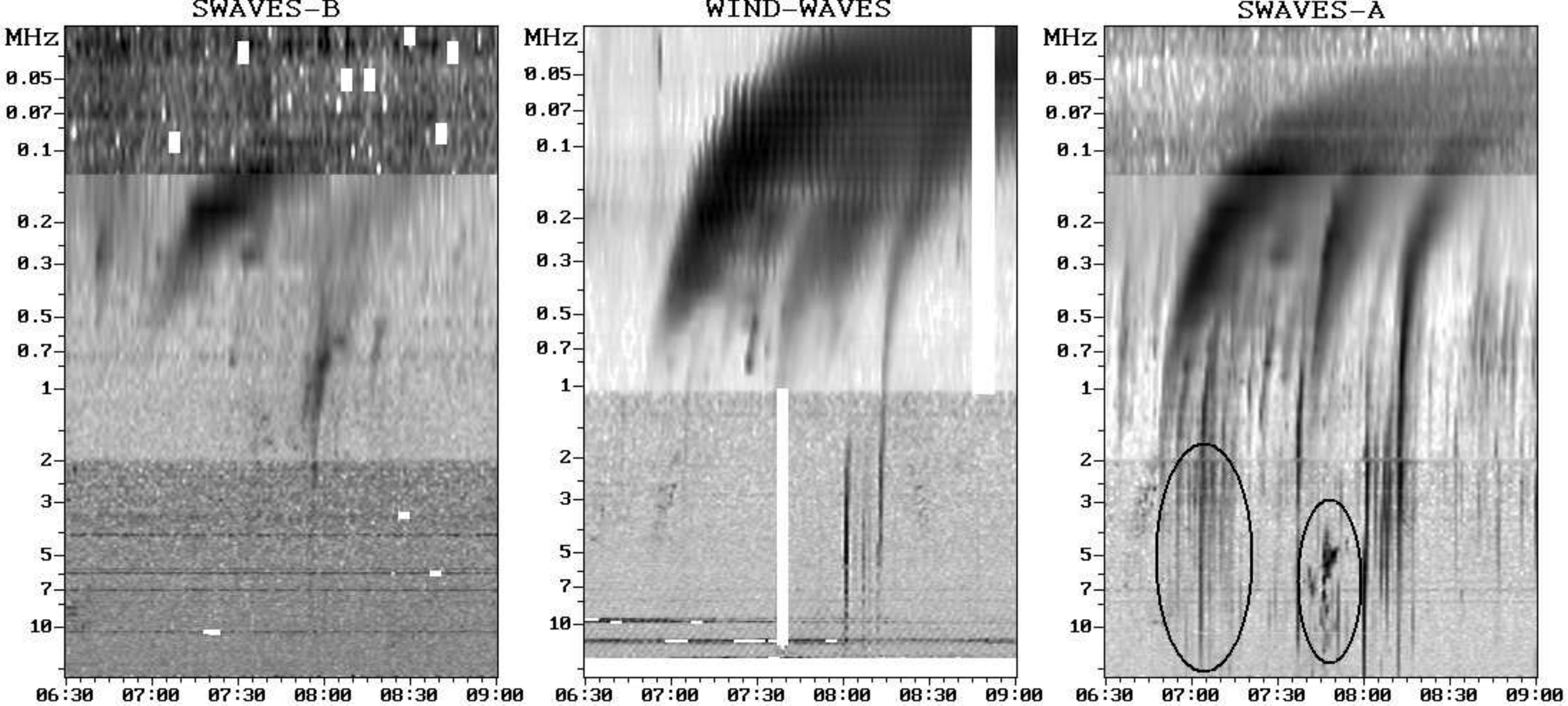}
\end{center}
\caption{Dynamic spectra from WAVES: STEREO-B (left), WIND (middle) and STEREO-A 
(right).}
\label{WAVES}
\end{figure}
apparently due to the occultation of the source by the solar disk; thus the only 
source of information is the SWAVES-A receiver aboard STEREO-A which, however, 
observes below 16\,MHz that corresponds to a radial distance $R\sim2.3$\,R\Sol, 
according to the Newkirk model. The dynamic spectrum (Figure \ref{WAVES}) shows a 
group of type III bursts from 06:48 to 07:18 UT; this is after the filament 
activation and nearly at the same time as the appearance of the first flare kernels 
in the STEREO 304 images (06:47 UT). The type IIIs were seen near the earth by 
WIND/WAVES at lower frequencies, as well as by SWAVES-A; they were detectable 
down to at least 30\,kHz, well into the interplanetary medium. 

A second instance of metric emission was from 07:35 to 07:54 UT, which was after 
the sequence of SSRT images shown in Figure \ref{SSRT}. This comprised a strong 
type III at 07:37 and a type II-like emission with fundamental-harmonic structure; 
the middle frequency of this feature was 5.51\,MHz at 07:48 UT and its logarithmic 
drift rate was $d\ln f/dt=-6.5\times10^{-4}$\,s$^{-1}$.

\begin{figure}
\begin{center}
\includegraphics[width=.8\textwidth]{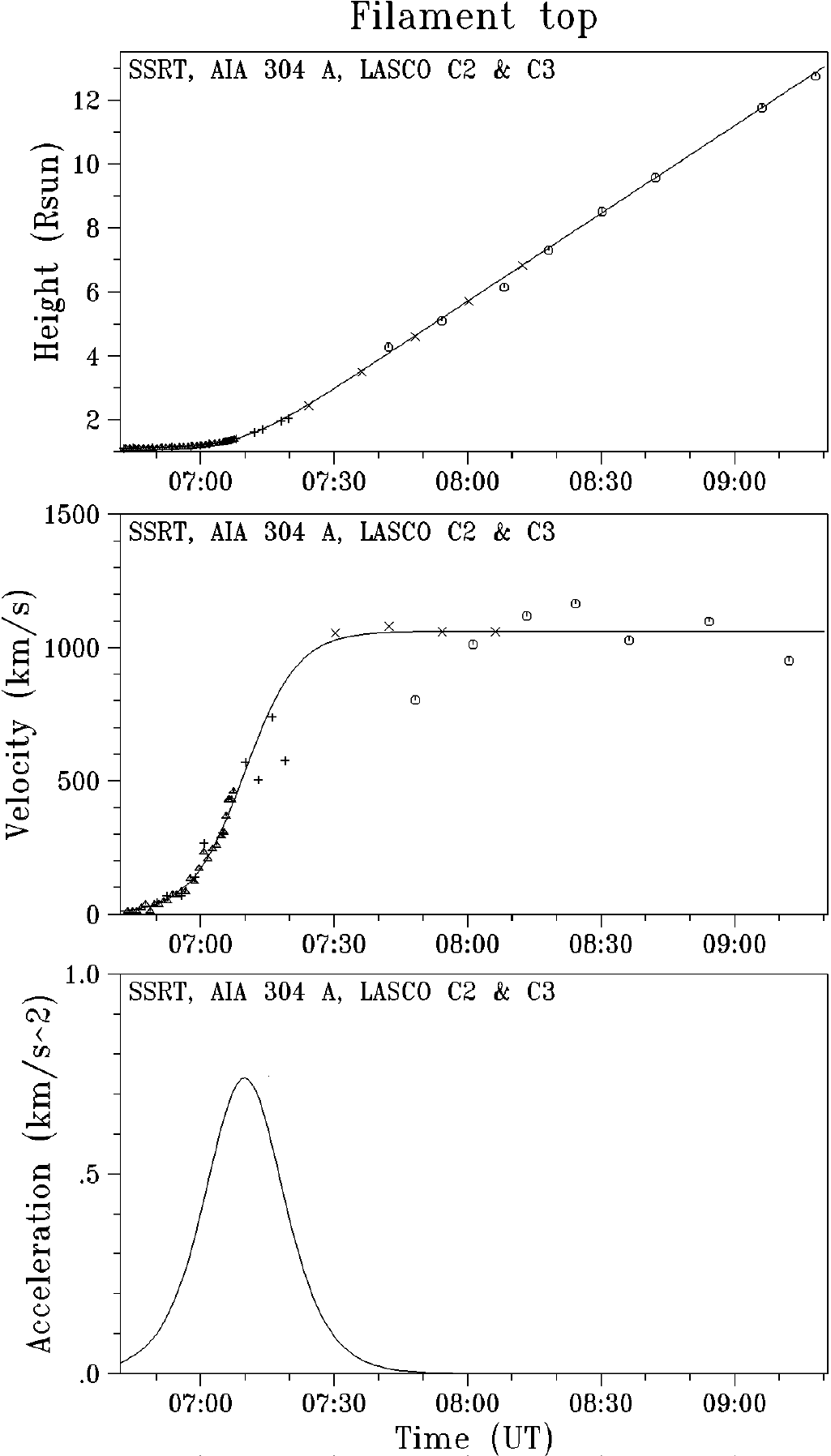}
\end{center}
\caption{Top: the height of the top of the filament as a function of time; 
triangles are from 304 \AA\ images, ``+'' from SSRT, ``x'' from LASCO 
C2, circles from LASCO C3. Middle: the corresponding velocity, computed 
numerically. The lines show the fit to the functions given in Equations (\ref{ht})
and (\ref{vel}). Bottom: the acceleration, computed from the fit.
}
\label{htvel}
\end{figure}

\section{Dynamics}
\label{dynamics}
The AIA/SDO data show a complex dynamic behavior of the eruption, in particular 
at the region of interaction of the two filaments as evidenced in the 
AIA 304\,\AA\ images in movies 1 and 2. We will outline here the motion of the 
main filament and we will defer a detailed study to a subsequent publication.

In spite of the complex dynamics in the interaction region, the main filament 
rises smoothly, as shown in the upper panel of Figure \ref{htvel}, which presents 
the projected height of the top of the filament as a function of time, measured 
in 304 \AA, SSRT and LASCO C2 and C3 images. The values among the four data sets 
are fairly consistent, with SSRT bridging the gap between AIA and LASCO data. The 
filament can be followed up to a height of $\sim7$\,R\Sol\ in the C2 field of 
view and up to $\sim13$\,R\Sol\ in C3. The velocity, shown in the middle panel of 
Figure \ref{htvel}, also shows a smooth rise and stabilizes to $\sim1100$\,km\,s$^{-1}$ 
after $\sim$07:30 UT. 

We fitted the projected height of the filament top, $h(t)$ with the smooth 
function:
\begin{eqnarray}
h(t)=h_{t_1}&+&\frac{v_f+v_o}{2}(t-t_1) \nonumber \\
            &+&\frac{v_f-v_o}{2}\,\tau\, \ln\, \cosh\left(\frac{t-t_1}{\tau}\right)
\label{ht}
\end{eqnarray}
proposed by Sheeley et al. (2007). Here $v_o$ and $v_f$ are the initial and 
final (asymptotic) values of the expansion velocity, $t_1$ is the time when the 
velocity attains its average value, $(v_f+v_o)/2$, $h_{t_1}$ is the corresponding 
height and $\tau$ is the time-scale of the acceleration. The corresponding 
velocity, $v(t)$, is:
\begin{equation} 
v(t) = \frac{v_f+v_o}{2} + \frac{v_f-v_o}{2} \tanh \left(\frac{t-t_1}{\tau}\right) 
\label{vel}
\end{equation}
and the acceleration, $a(t)$ is:
\begin{equation}
a(t) = \frac{v_f-v_o}{2\tau}\left[1-\tanh^2 \left(\frac{t-t_1}{\tau}\right)\right] 
\label{acc}
\end{equation}
The acceleration maximizes at $t=t_1$ with a maximum value of
\begin{equation}
a_{max}=\frac{v_f-v_o}{2\tau}
\label{amax}
\end{equation}
the width of the $a(t)$ curve being approximately $2\tau$.

We fitted both the projected height to Equation (\ref{ht}) and the velocity, 
computed numerically as the forward difference of consecutive heights, to 
Equation (\ref{vel}). The numerical computation of the acceleration was much too 
noisy to be used for a fit to Equation (\ref{acc}). The obtained parameters from 
the two fits had similar values, so we used their average to plot the smooth 
curves in Figure \ref{htvel}. The acceleration derived from the fit (lower panel 
of Figure \ref{htvel}) has a maximum value of 0.74\,km\,s$^{-2}$ around 07:09:53 
UT, and a width of $2\tau\approx 24$\,min.

\begin{figure}[ht]
\begin{center}
\includegraphics[width=.8\textwidth]{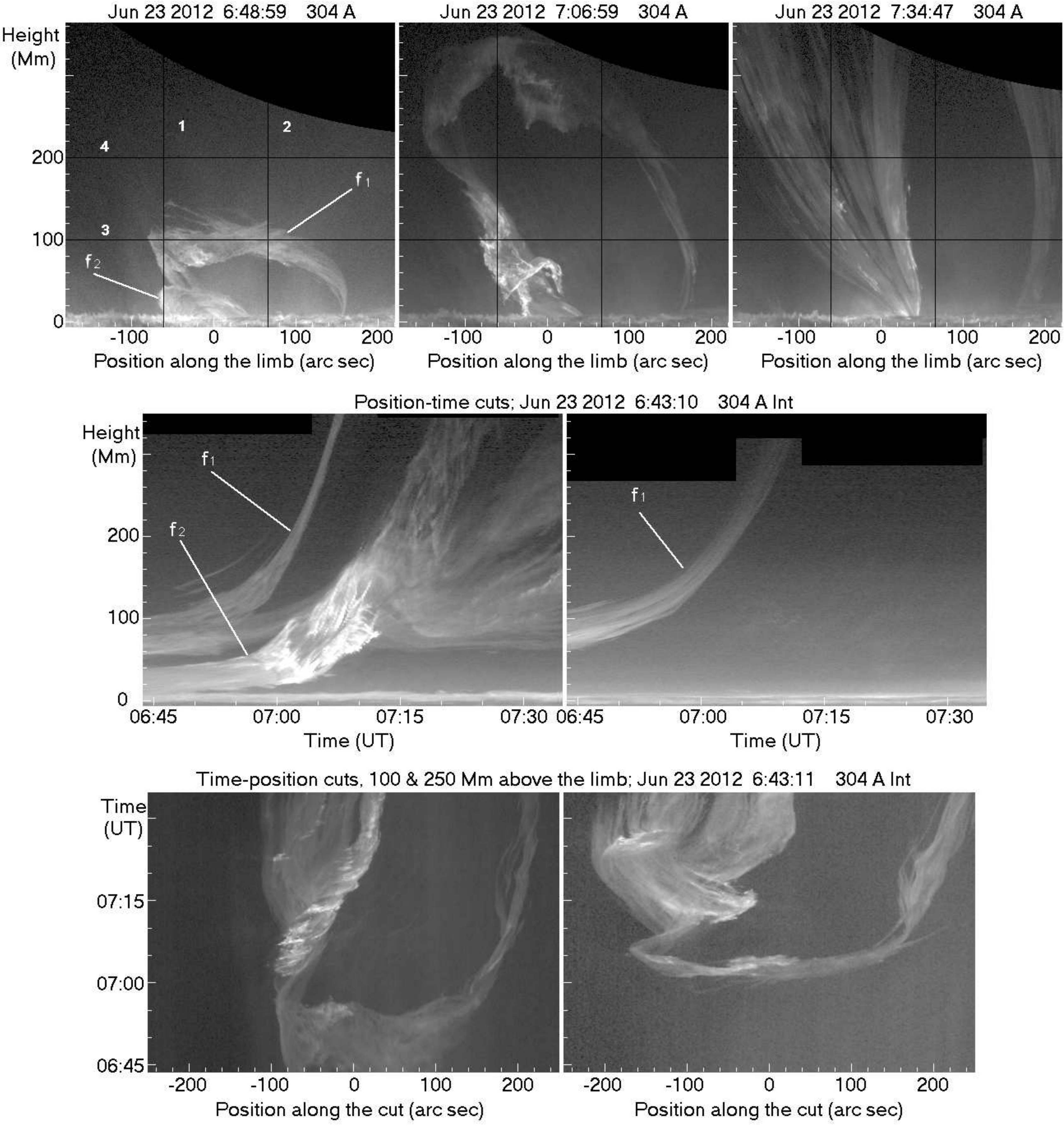}
\end{center}
\caption{Top row: 304 \AA\ images at three instances during the filament eruption. 
The images have been re-mapped in a coordinate system which has distance along the 
limb as the x-axis and height above the limb as the y-axis. Middle row: Intensity 
as a function of time and position along radial cuts marked 
1 and 2 on the top row. Bottom row: Intensity as a function of position and time 
at 100 and 200\,Mm above the limb, along cuts 3 and 4.}
\label{cuts}
\end{figure}

The dynamics is much more complex in the region of interaction of the two 
filaments. We show in the top row of Figure \ref{cuts} three 304\,\AA\ AIA images 
at characteristic phases of the eruption: before the 2-filament interaction, during 
and after. In the middle row we present images of the intensity as a function of 
time and position along the lines marked 1 and 2 in the upper row; they show 
very well the rise of the two filaments, f$_1$ and f$_2$. Cut 1 shows 
clearly that the phase of rapid rise of f$_2$ was delayed by a few minutes with 
respect to f$_1$; it also shows very well the interaction, which appears as 
bright emission between 06:55 and 07:13 UT. Note also that the interaction 
appears to have no effect to the top of the main filament (f$_1$), as evidenced 
by cut 2. 

The cuts parallel to the limb (bottom row of Figure \ref{cuts}) also show very 
well the interaction region. The first thing to notice is that the motion is much 
more complex in the north leg of the filament, where the interaction occurred, 
than in the south. Strong twisting motions are observed and the material spreads 
out considerably, in contrast to the south leg where twisting motions are weak and 
the material remains fairly confined.

\section{Contribution of High Temperature Plasma to the microwave emission}
The EUV bands of AIA, with the exception of the one at 304\,\AA, can be used for 
computations of the differential emission measure (DEM) in the temperature range
between $\log T$ of 5.5 to 7.5. The DEM, in turn, can be used to compute the 
contribution of the plasma within this temperature range to the microwave 
brightness temperature, $T_b$. Obviously, as discussed in section \ref{over}, 
there is significant contribution to the microwave emission from plasma below this 
temperature range, nevertheless the value computed from the DEM can be considered 
as a lower limit.

The microwave brightness temperature is given by
\begin{equation}
T_b=\int_0^\tau T_e e^{-\tau} d\tau \label{tb}
\end{equation}
where $T_e$ is the electron temperature and $\tau$ the optical depth along the line 
of sight. The optical depth is determined by the electron temperature and the 
electron density, $N_e$:
\begin{equation}
d\tau = -\frac{\xi}{f^2}\frac{N_e^2}{T_e^{3/2}}\,d\ell \label{tau}
\end{equation}
where $f$ is the observing frequency, $\ell$ is the distance along the line of 
sight and $\xi$ is a slowly varying function of temperature and density. We have 
taken $\xi=0.2$, taking into account the effect of He and heavy elements (Chambe 
and Lantos, 1971).

Suppose now that the electron temperature varies monotonically along the line of 
sight, as, {\it e.g.}, in the case of a stratified atmosphere; then the integration 
of Equations (\ref{tau}) and (\ref{tb}) can be done with $T_e$ as a variable rather than $\ell$ or 
$\tau$. The optical depth can be expressed as a function of $T_e$ as follows:
\begin{equation}
d\tau = -\frac{\xi}{f^2}\frac{N_e^2}{T_e^{3/2}}\,\frac{d\ell}{dT_e}\,dT_e=
        -\frac{\xi}{f^2}\frac{\varphi(T_e)}{T_e^{3/2}}\,dT_e \label{tau2}
\end{equation}
where
\begin{equation}
\varphi(T_e)=N_e^2\,\frac{d\ell}{dT_e} \label{phi}
\end{equation}
is the DEM. This integrates to:
\begin{equation}
\tau(T_e) = -\frac{\xi}{f^2}\int_{T_{e_1}}^{T_{e}} \frac{\varphi(T_e)}{T_e^{3/2}}\,dT_e
\label{tau3}
\end{equation}
where $T_{e_1}$ is the electron temperature nearest the observer. Using $T_e$ as a
variable, Equation (\ref{tb}) gives for the brightness temperature:
\begin{equation}
T_b = \frac{\xi}{f^2}\int_{T_{e_1}}^{T_{e_2}} e^{-\tau(T_e)} \frac{\varphi(T_e)}{T_e^{1/2}}\,dT_e
\label{tb2}
\end{equation}
where $T_{e_2}$ is the electron temperature further away from the observer.

Note that expressions (\ref{tau2}) to (\ref{tau3}) are valid even if the temperature variation along 
the line of sight is not monotonic. This is not the case for (\ref{tb2}), due to 
the $e^{-\tau}$ factor; however, Equation (\ref{tb2}) is still valid in the 
optically thin case, because then the exact location of each temperature layer is 
not important.

To calculate DEMs from the AIA observations we used the method described in 
Plowman et al. (2013), hereafter PKM13. The input to the PKM13 
method is the observed AIA intensity in its 6 coronal channels centered around 
94, 131, 171, 193, 211 and 335 \AA, which cover the range 5.5-7.5 in logT. PKM13  
is a very fast iterative method based  on regularization. The temperature 
response functions of the AIA channels are used as basis functions in the 
inversion. For more details on the method and its performance we refer the reader 
to the PKM13 article. Other methods for DEM determinations from AIA data are 
described by Aschwanden et al. (2013) and in Hannah \& Kontar (2012). Note, 
that it is well-known that the DEM inversion problem is an ill-posed one, i.e., 
there are more unknowns than knowns (e.g. Craig \& Brown, 1976) and thus 
the solutions we (or anybody) calculate are not unique. Morever, the PKM13 method 
gives occasionally negative values for the DEM; these were ignored in our further 
analysis.

For the computation we used the AIA observations around 07:02 and 07:32 UT. 
The AIA level-1 images were first co-registered to a common center and binned to 
the same pixel size using the aia\_prep.pro routine. In the DEM calculations we 
used the most recent available AIA temperature response functions (V4) which 
take into account an inter-calibration between AIA full disk intensities and 
SDO/EVE spectral irradiance data as well as a fix to the temperature response 
function of mainly the 94\,\AA\ channel for which the CHIANTI database used in 
the calculation of the AIA temperature functions lacks several spectral lines. 
Our calculations were carried out over a sub-field around the erupting filament.

One way to access the success of the performed DEM inversions is to compare the 
AIA intensities ($I_{DEM}$) predicted from the calculated DEMs with the observed 
intensities ($I_{obs}$). We thus defined the normalized intensity residual as
\[R_{norm}=\frac{I_{obs}-I_{DEM}}{I_{obs}}\] and calculated it for all 6 AIA 
channels used in the DEM determination. The above quantity for most channels 
exhibits a well-defined peak in the range  of $\approx$ [-0.1, 0.1] with FWHM 
widths in the range of $\approx$  [0.05, 0.4]. $R_{norm}$ for the 94\,\AA\ band 
shows a broader distribution of residuals, in the range [-0.4, 0.4]. From 
the above we conclude that the calculated DEMs do a rather good job in 
reproducing the observed intensities to within 5 to 40 $\%$ which is deemed 
acceptable given all the uncertainties and assumptions entering into the DEM and 
temperature response function determinations.

As a further check of our computations, we used our results to calculate the 
total volume emission measure and compared it to the one deduced from the ratio of 
the two soft X-ray GOES channels. At 07:32 UT we got a value of 2.11$\times10^{48}$\,cm$^{-3}$ 
from our measurements and 2.0$\times10^{48}$\,cm$^{-3}$ from GOES, which also gave 
an average temperature of $\sim10^7$\,K; the proximity of the DEM values is 
striking so is the consistency of the temperature (see next paragraph).
 
\begin{figure}
\begin{center}
\includegraphics[width=.8\textwidth]{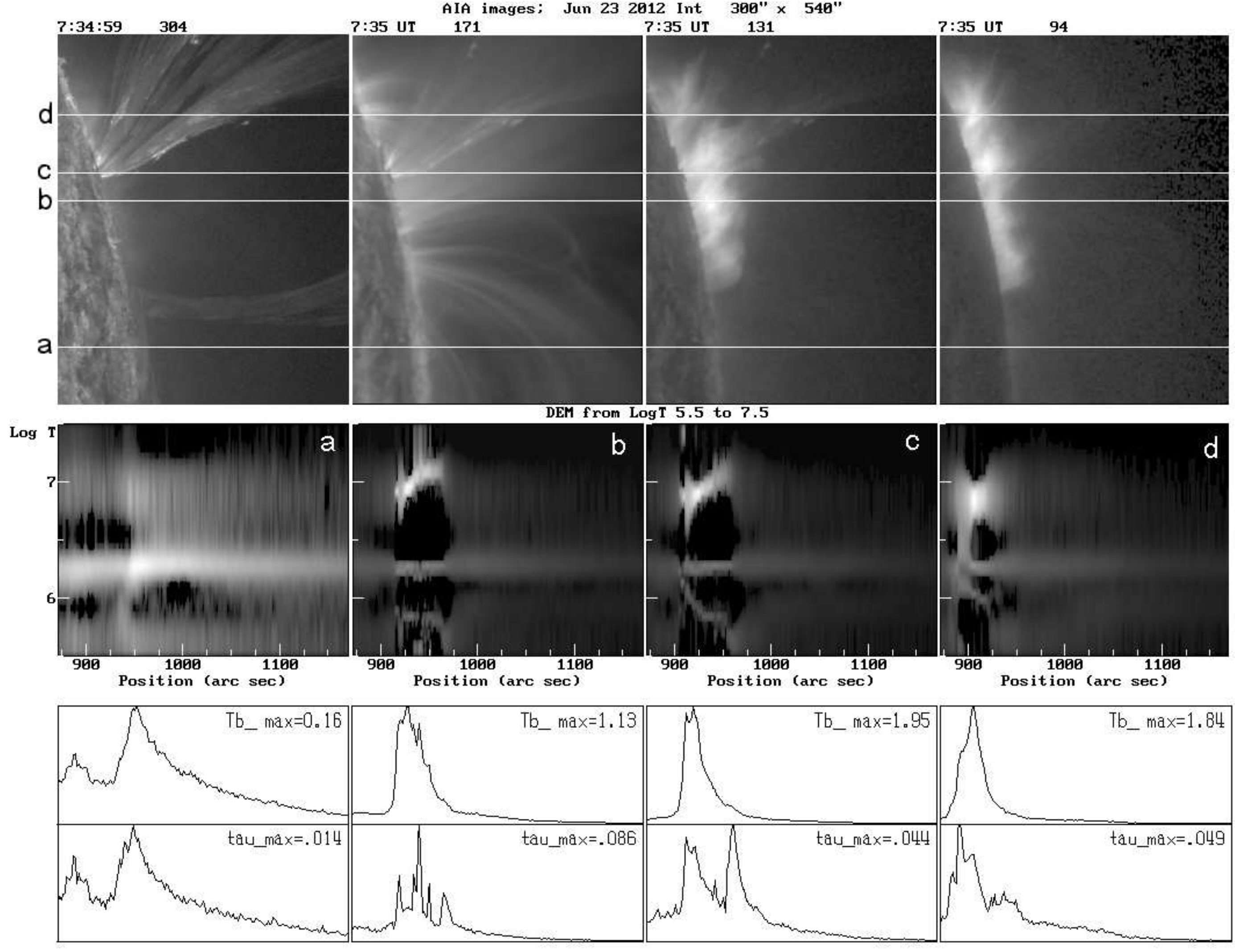}
\end{center}
\caption{Computation of differential emission measure, $\tau$ and $T_b$. Top : 
AIA images in the 304, 171, 131 and 94 \AA\ bands. Middle: The computed DEM as a 
function of position and $\log T$, along the cuts a, b, c and d marked in the top 
panel; the position is measured from the center of the disk. Bottom:
The corresponding optical depth and brightness temperature at $\lambda=5.2$\,cm;
the values of $T_{b_{max}}$ are in $10^5$\,K.}
\label{dem}
\end{figure}

Figure \ref{dem} shows examples of our computations using AIA images near 07:35 
UT. The middle row shows the DEM as a function of position and $\log T$, along the 
four cuts marked $a$ to $d$ in the images of the upper panel. Here the 304 \AA\ image
is given for reference, since it was not used in the DEM computation; the other 
images sample low and high temperature regions. The bottom plots show the optical 
depth and the brightness temperature at the SSRT frequency, computed as 
outlined above, under the optically thin approximation. Note that in the bright 
region above the limb most of the plasma is near $\sim10^7$\,K, which is 
consistent with the GOES results.

Let us first note that in all cases the optical depth is much less than unity, so 
that the optically thin approximation is valid. Cut $a$ represents a very quiet 
coronal region and the corresponding DEM shows a peak around $\log T_e=6.25$, 
which corresponds to the coronal temperature in the quiet sun. The maximum 
brightness temperature is only 16\,000\,K, with the peak right after the limb, as 
expected. Cuts $b$ and $c$, that cross regions near the limb which appear bright 
in both the 131 and the 94 \AA\ bands (section \ref{over}), show a clear high 
temperature component near $\log T_E \sim 7$ and the brightness temperature goes 
up to $\sim 2\times10^5$. The high temperature component has a slightly smaller 
value in cut $d$, in which the 131 \AA\ emission is weaker. In all cases there is 
a weak component at $\log T_e < 6$, apparently corresponding to the upper Transition Region.

\begin{figure}[h]
\begin{center}
\includegraphics[width=.8\textwidth]{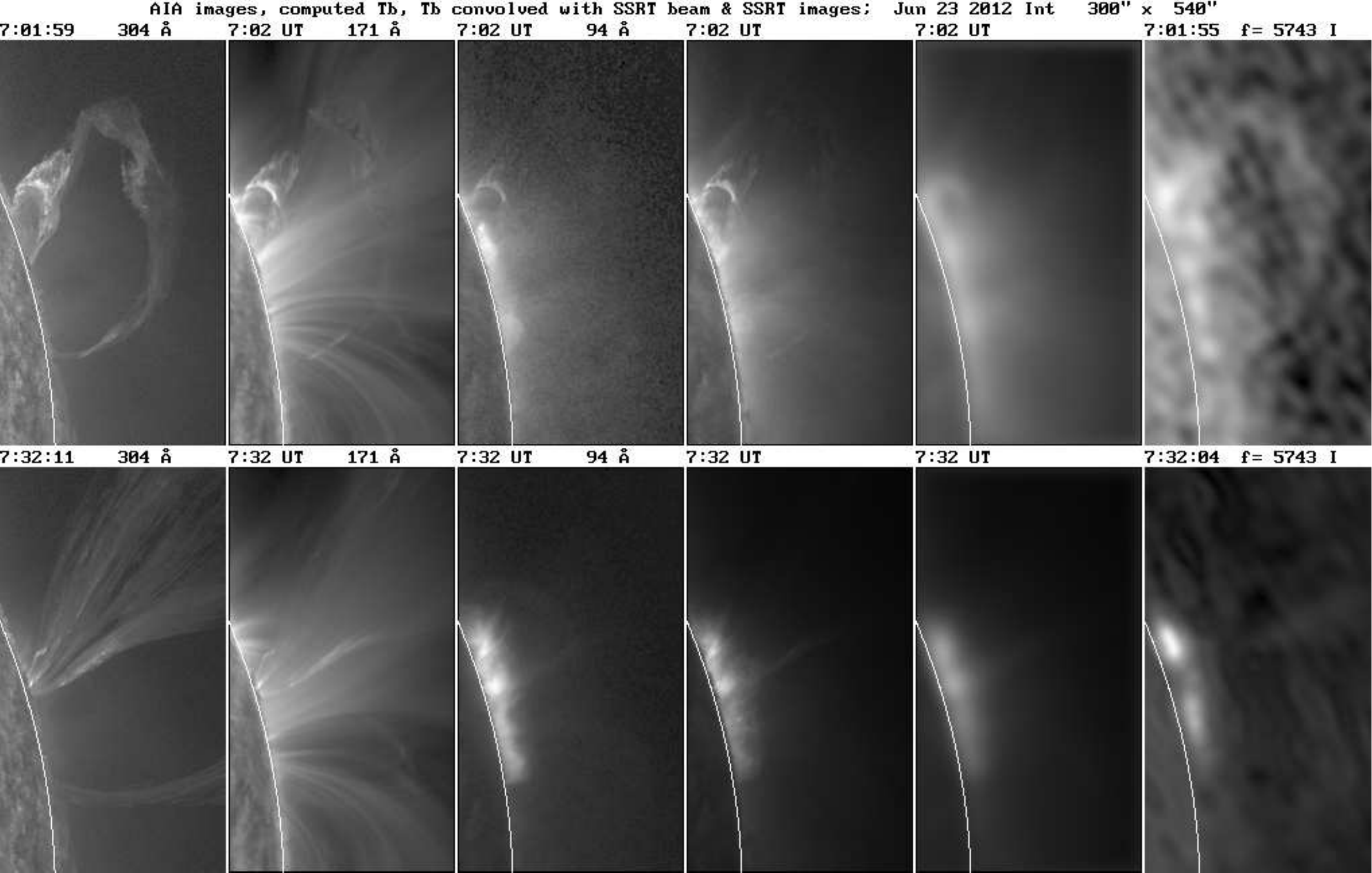}
\end{center}
\caption{Comparison of the computed brightness temperature with the SSRT images at 
two instances, 07:02 UT (top panel) and 07:32 UT (bottom panel). The first three
columns show AIA images at 304, 171 and 94 \AA\ for reference. The forth column 
is the brightness temperature computed from the DEM, the fifth is the same after 
convolution with the 23\arcsec\ SSRT beam. The last column shows the corresponding 
SSRT images, displayed in the same intensity scale.}
\label{comp}
\end{figure}

We will now compare the computed brightness with the SSRT images. This is done in 
Figure \ref{comp}, where we selected two instances: one at 07:02 UT, where the main 
filament is still in the AIA field of view and another at 07:32 UT, where the bright 
region next to the limb is well visible. There is no trace of the filament itself 
in the computed $T_b$, which verifies that its temperature is much below the range 
of the DEM computed from the AIA images. What is clearly visible is the region of 
interaction of the two filaments and the bright region just above the limb. Note,
however, that the observed spatial distribution of the latter is not the same as the
computed; in particular, the SSRT shows two bright blobs, while the computed 
$T_b$ is more evenly distributed along the limb.

Comparing the computed brightness temperature with the observed, we note that the 
maximum computed $T_B$ at 07:32 UT is $3\times10^5$\,K and drops to $1.6\times10^5$\,K 
after the convolution with the SSRT beam; this is 65\% of the observed value of 
$2.5\times10^5$\,K. 

\begin{figure}[h]
\begin{center}
\includegraphics[width=.8\textwidth]{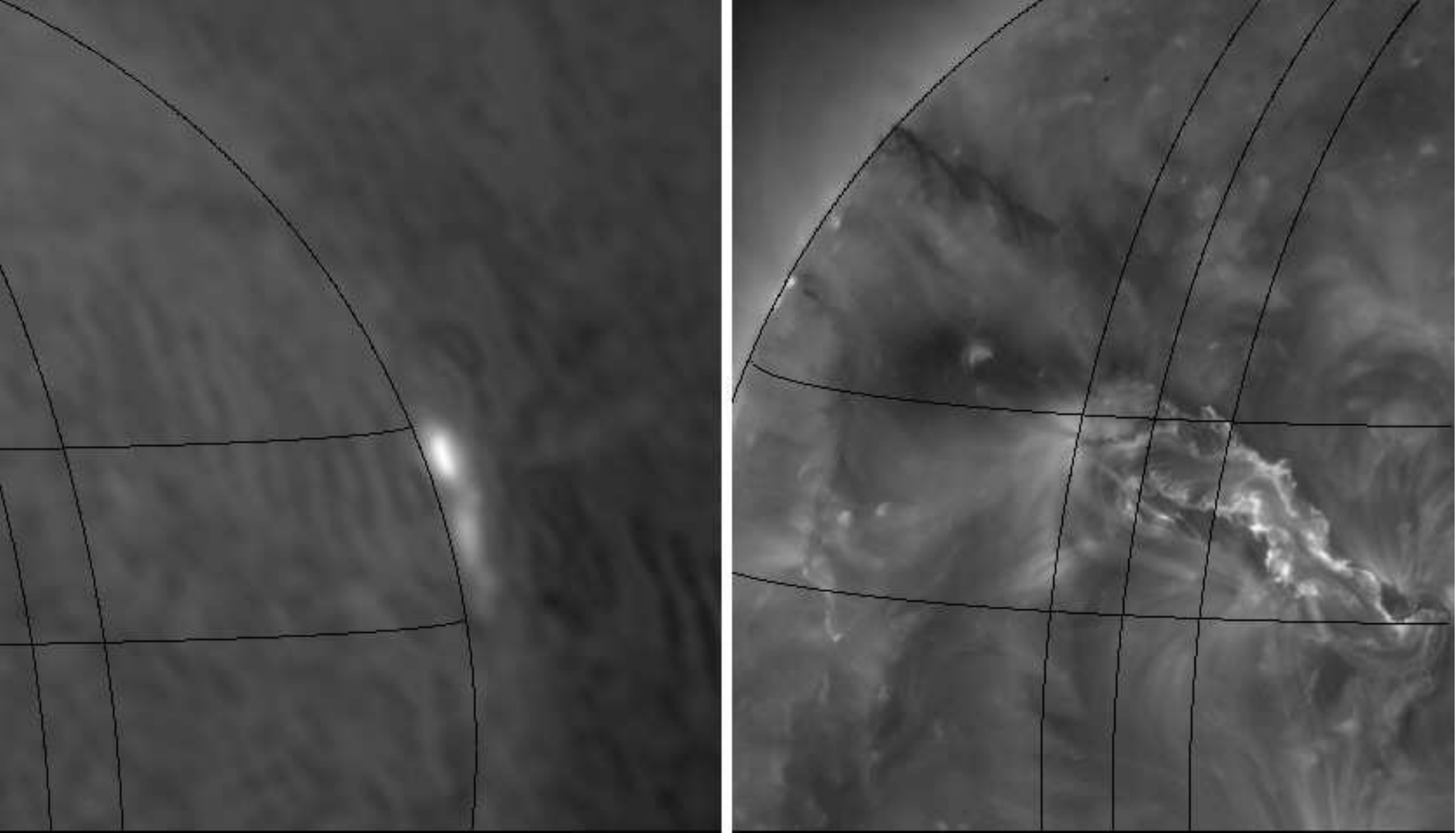}
\end{center}
\caption{SSRT image at 07:35 UT (left) and STEREO-A 195\,\AA\ image at 07:36 UT. 
Lines are drawn at latitudes of 7 and 23 degrees; In each image the position of 
the limb of the other image is marked, together with the position of points 
5\arcsec\ in front and 5\arcsec\ behind the limb.}
\label{brlimb}
\end{figure}

Let us now consider the origin of the bright region above the limb that developed 
at the late stage of the eruption. From its geometry it is obvious that it comes 
from behind the limb and one possible origin is the tops of hot flare loops. Figure
\ref{brlimb} shows an SSRT image and the corresponding STRERO-A image at 195\,\AA. 
The comparison of the two shows that the northern bright blob is located between
the flare ribbons in the E part of the flare, while the southern blob is between  
the ribbons in the W part of the flare; thus they might indeed be manifestations 
of hot flare loop tops.

\begin{figure}[h]
\begin{center}
\includegraphics[width=.8\textwidth]{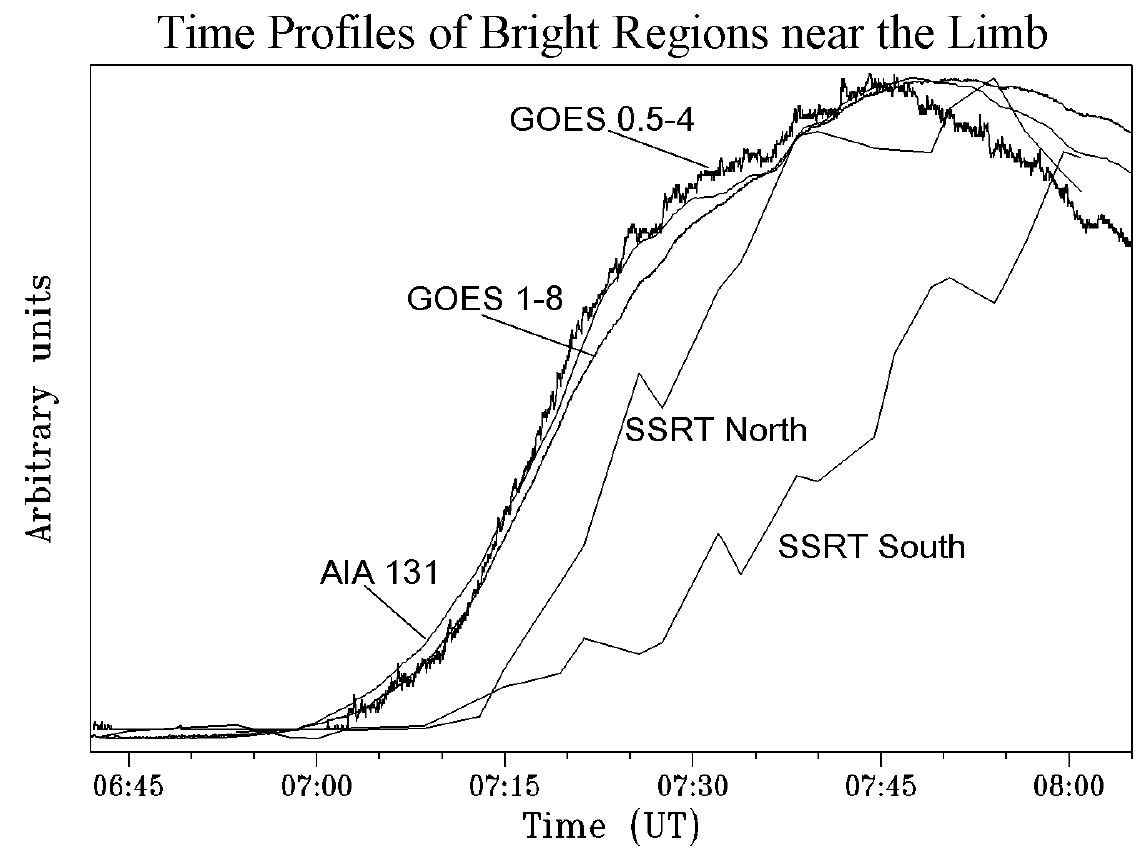}
\end{center}
\caption{Normalized time profiles of the bright region above the limb in GOES, 
AIA 131\AA\ and SSRT.}
\label{brprof}
\end{figure}

Further information about the origin of the emission can be obtained from its time 
profile (Figure \ref{brprof}). The form of the profiles in the two GOES channels 
is practically identical during the rise, which indicates that there is no 
hard x-ray component in the emission, apparently because any such component is 
occulted. Unfortunately, during the event there are no RHESSI data that would 
confirm the absence of hard X-ray emission. The time profile in the AIA 131\,\AA\ 
band follows closely the GOES profiles. The SSRT shows two distinct regions 
({\it cf} Figure \ref{brlimb}), with the one in the north brightening first. 

Both bright regions show too slow a rise to be anything but of thermal origin, 
although the northern region rises faster than GOES. It is interesting, moreover, 
that the rise of the SSRT emission is delayed by $\sim$4\,min with respect to 
GOES. One possible explanation is in terms of optically thin emission from a 
cooling hot plasma, since $T_b\propto T_e^{-1/2}$ in this case. The maximum of the 
northern SSRT region occurs near the peak of the GOES profiles, while the 
southern region, which is originally weaker than the northern, attains its 
maximum about 10 minutes later. If we accept the interpretation given previously 
that the origin of the bright regions is the top of hot flare loops, the delay of 
the southern region could be explained in terms of geometric effects, since 
its origin is located further away from the solar limb as seen from the earth. 

\begin{figure}[h]
\begin{center}
\includegraphics[width=.8\textwidth]{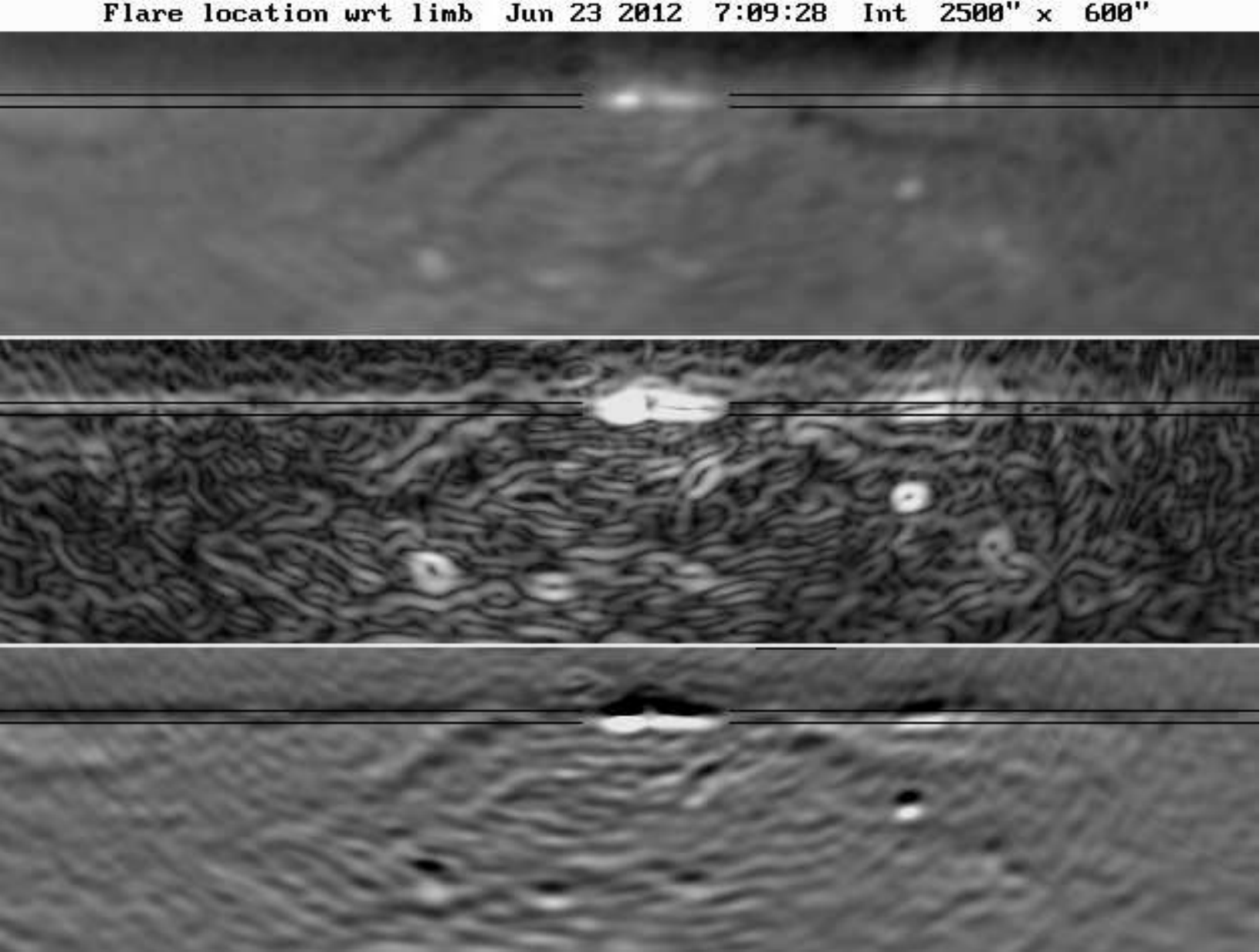}
\end{center}
\caption{SSRT image at 07:09:28, re-mapped in a coordinate system along and 
perpendicular to the limb (top). The lower black horizontal line marks the 
photospheric limb and the upper the radio limb; the latter appears as a bright 
band in the gradient of the image (middle panel) and as a dark band in the radial 
derivative (lower panel).}
\label{limb}
\end{figure}

Another interesting result is that the SSRT emission is below the radio limb 
(Figure \ref{limb}). As the AIA images certify that the source of the emission is 
behind the limb, this implies that the region between the optical and the radio 
limb is optically thin at 5.2\,cm.

\section{Summary and conclusions}
The combination of the SSRT observations with data from SDO and STEREO give a more 
complete view of the filament eruption and the associated CME that was observed 
above the W limb on June 23, 2012. This is an important improvement compared to 
previous works that used EIT for comparison (Uralov et al. (2002), 
Grechnev et al. (2006)). The SSRT detected the filament up to a height 
of more than 1\,R\Sol\ above the limb, at a time that the CME was visible in the 
LASCO C2 coronograph. This made possible to verify that the filament coincides 
with  the core of the CME, as already pointed out by Grechnev et al. (2006). 
Thus the SSRT bridges the gap between the field of view of AIA/SDO and the 
occulting disk of C2.

The ascent of the filament could be followed up to $\sim13$\,R\Sol\ in the C3 
coronograph. Its accelerating phase lasted for about 50 min and after that its 
projected velocity stabilized to $\sim1100$\,km\,s$^{-1}$, a phase that lasted 
as long as the filament was detectable in C3 ($\sim110$\,min). Fitting the 
data to the smooth curve proposed by Sheeley et. al. (2007), we deduced a 
maximum accelaration of 0.74\,km\,s$^{-2}$, attained around 07:09:53 UT. During 
the rise of the acceleration the first flare kernel at 304\,\AA\ was recorded in 
STEREO-A images (at 07:47) and a prominent group of type III burst appeared in the 
dynamic spectrum of STEREO-A/WAVES (from 06:48 to 07:18 UT).

Later on, from 07:40 to 07:54 UT, a feature resembling a short type II burst was 
detected in the STEREO-A/WAVES dynamic spectrum; at 07:47 UT it was at 5.51\,MHz 
and had a frequency drift $d\ln f/dt=-6.5\times10^{-4}$\,s$^{-1}$. At that time 
the projected height of the filament was 4.6\,R\Sol\ and its projected velocity
1060\, km\,s$^{-1}$. 

The emission frequency of the type II-like feature 
corresponds to an electron density of $3.77\times10^5$\,cm$^{-3}$. According to 
the Newkirk (1961) model, this density occurs at a height of 
4.5\,R\Sol\ which is comparable to the observed, 4.6\,R\Sol; by contrast, the 
Saito (1970) model gives a height of only 3\,R\Sol, well below the observed. The 
observed frequency drift corresponds to a radial velocity of 1800\,km\,s$^{-1}$ in 
the Newkirk (1961) model and 600\,km\,s$^{-1}$ in the Saito (1970) 
model, compared to the observed speed on the plane of the sky of 1060\,km\,s$^{-1}$.
If the proximity of the observed height to that deduced from the Newkirk model 
signifies that the prominence was practically on the plane of the sky, the 
difference between the model radial velocity and the measured projected velocity 
might be attributed to the inclination of the velocity vector with respect to the 
sky plane.

Both AIA/SDO and STEREO data showed the interaction of the main filament, 
initially located above the neutral line of AR 11506, with a smaller filament, 
located almost in the perpendicular direction. A similar two-filament interaction 
was reported by Uralov et. al. (2002), who considered it as the origin of 
the eruption. In our case the interaction, which led to intense emission in the 
high temperature AIA bands, as well as to strong twisting motions in the northern 
leg of the filament, occurred at a time when both filaments were already rising  
with increasing acceleration. It is thus rather unlikely that the filament 
interaction initiated the eruption, although it must have had a significant 
contribution to the acceleration, the maximum of which occurred after the 
interaction.

Morphologically the filament was very similar in the SSRT images, in H$\alpha$, 
and in the AIA 304 and 1600\,\AA\ bands. A closer examination and the comparison 
with the AIA images revealed three components in the microwave emission: a low 
temperature component, associated with the filament proper, a high temperature 
component associated with the filament interaction region and an even higher 
temperature component associated with the top of hot post-flare loops. Thus the 
SSRT images probe efficiently a very wide temperature range, from $\sim10^4$ to 
$\sim10^7$\,K.

We found no evidence of microwave non-thermal emission, which is natural due 
to the fact that the flare footpoints were behind the limb. We thus computed the 
brightness temperature of the thermal emission from the differential emission 
measure in the range of $\log T$ between 5.5 and 7.5, deduced from the AIA high 
temperature bands. The DEM computed in this way was consistent within 5\% with the
volume emission measure computed from the GOES data. For the bright regions 
associated with the top of hot post-flare loops we found that the computed peak 
brightness temperature, after convolution with the SSRT beam, was 35\% lower than 
the observed.

Several works in the past have shown inconsistencies between the radio emission 
computed from EUV data and the observations. In the quiet sun, for example, in order 
to match the computations with the observation one has to take models of the cell 
interior rather than of the average quiet sun (see, {\it e.g.\/}, the review by
Shibasaki et al., 2011). In another work Zhang et al. (2001), using a 
three temperature EUV model found that the computed $T_b$ was two times greater 
than the observed. Using simultaneous WRST and HXIS data, 
Alissandrakis et al. (1988) found very little contribution of thermal emission to 
the total radio emission in the post flare phase (which is considered to be of 
thermal origin) for most of the events they studied. However, 
Grechnev et. al. (2006) got a fairly good match between the flux observed by 
the SSRT and that computed on the basis of GOES data.

In our case the difference between the computed and observed $T_b$ can be 
attributed to several causes. One explanation is that the missing part of the 
observed emission comes from plasma at temperatures lower that $3.1\times10^5$\/K, 
which are below the range of the AIA-computed DEM. Another possibility is that 
the difference is due to calibration inaccuracies. A third possibility is the 
presence of a non-thermal component which, in the absence of hard X-ray data, 
cannot be excluded.

Finally, we noticed that the loop-top associated emission in the SSRT images came 
from a region located below the radio limb; this implies that the region between 
the optical and the radio limb is optically thin at 5.2\,cm. 

An extension of the present work will be a detailed  study of the possible 
triggering and driving mechanism(s) of the filament eruption. For example, the 
study of the temporal associations between the flare emissions (AIA, GOES and 
EUVI light-curves) and the filament acceleration will possibly supply hints on 
whether the eruption was triggered by an ideal or non-ideal process ({\it e.g.\/},
Bein et. al., 2012). Moreover, it will be interesting to study 
characteristics of the eruption such as the evolution of its rotation 
and twist, to see if they comply with predictions of CME models ({\it e.g.\/},
Lynch et al. (2009), Kliem et al. (2012)).

\vspace{0.5cm}
\small
This work benefited greatly from the data available in the AIA/SDO, 
SECCHI/STEREO, LASCO/SOHO, GOES, GONG, WIND/WAVES and STERO/WAVES data bases. The 
authors would like to thank all those who worked for the development of these 
instruments, for their operation and for making public the data. The AIA data 
used here are courtesy of SDO (NASA) and the AIA consortium. The SECCHI data are 
courtesy of STEREO and the SECCHI consortium. SOHO is an international 
collaboration between NASA and ESA. LASCO was constructed by a consortium of 
institutions: the Naval Research Laboratory (Washington, DC, USA), the 
Max-Planck-Institut fur Aeronomie (Katlenburg- Lindau, Germany), the Laboratoire 
d'Astronomie Spatiale (Marseille, France), and the University of Birmingham 
(Birmingham, UK). The Global Oscillation Network Group (GONG) project, is managed 
by the National Solar Observatory, which is operated by AURA, Inc. under a 
cooperative agreement with the National Science Foundation. The STEREO/WAVES and 
Wind/WAVES experiments are a collaboration of NASA Goddard Space Science Fight 
Center, the Observatory of Paris-Meudon, the University of Minnesota  and the 
University of California, Berkley.
 
The work in the Institute of Solar Terrestrial Physics is supported by the Ministry 
of Education and Science of the Russian Federation (projects 8407 and 
14.518.11.7047) and by the grants of RFBR (12-02-91161-GFEN-a, 12-02-00616, 
12-02-00173-a, 12-02-33110-mol-a-ved, and  12-02-31746-mol-a) and by a Marie 
Curie International Research Staff Exchange 
Scheme Fellowship within the 7th European Community Framework Programme.

S. Patsourakos acknowledges support from an FP7 Marie Curie Re-integration Grant 
(FP7-PEOPLE-2010-RG/268288) and from European Union (European Social Fund-ESF) 
and Greek national funds through the Operational Program "Education and Lifelong 
Learning" of the National Strategic Reference Framework (NSRF) - Research Funding 
Program: Thales. Investing in knowledge society through the European Social Fund.

C. Alissandrakis would like to thank the colleagues from ISTP for their warm 
hospitality during his stay there; he also thanks the Organizing Committee of the 
SPRO2012 conference held in Nagoya in November 2012 for funding his participation.

Finally the authors wish to thank H. Warren and P. Boerner for their help with the AIA 
temperature response functions.

\normalsize


\setlength{\parindent}{0cm}
\bigskip\bigskip
\Large{\bf{References}}
\normalsize

Alissandrakis, C.~E., Schadee, A., \& Kundu, M.~R.\ 1988, \aap, 195, 290 

Aschwanden, M.~J., Boerner, P., Schrijver, C.~J., \& Malanushenko, A.\ 2013, 
Sol. Phys., 283, 5 

Bein, B.~M., Berkebile-Stoiser, S., Veronig, A.~M., Temmer, M., 
\& Vr{\v s}nak, B.\ 2012, \apj, 755, 44 

Chambe, G., \& Lantos, P.\ 1971, Sol. Phys., 17, 97 

Chiuderi Drago, F., Alissandrakis, C.~E., Bastian, T., Bocchialini, K., 
\& Harrison, R.~A.\ 2001, Sol. Phys., 199, 115 

Craig, I.~J.~D., \& Brown, J.~C.\ 1976, \aap, 49, 239 

D'Azambuja, L.\ 1925, L'Astronomie, 39, 209 

Forbes, T.~G.\ 2000, \jgr, 105, 23153 

Gopalswamy, N.\ 1999, Proceedings of the Nobeyama Symposium, NRO Report 479, 141 

Gopalswamy, N., Shimojo, M., Lu, W., et al.\ 2003, \apj, 586, 562 

Grechnev, V.~V., Lesovoi, S.~V., Smolkov, G.~Y., et al.\ 2003, Sol. Phys., 216, 239 

Grechnev, V.~V., Uralov, A.~M., Zandanov, V.~G., Baranov, N.~Y., 
\& Shibasaki, K.\ 2006, \pasj, 58, 69 

Hannah, I.~G., \& Kontar, E.~P.\ 2012, \aap, 539, A146 

Kliem, B., T{\"o}r{\"o}k, T., \& Thompson, W.~T.\ 2012, Sol. Phys., 281, 137 

Klimchuk, J.~A.\ 2001, Washington DC American Geophysical Union Geophysical 
Monograph Series, 125, 143 

Kundu, M.~R.\ 1972, Sol. Phys., 25, 108 

Landi, E., Raymond, J.~C., Miralles, M.~P., \& Hara, H.\ 2010, \apj, 711, 75 

Lantos, P., Alissandrakis, C.~E., Gergely, T., 
\& Kundu, M.~R.\ 1987, Sol. Phys., 112, 325 

Lynch, B.~J., Antiochos, S.~K., Li, Y., Luhmann, J.~G., \& DeVore, C.~R.\ 2009, 
\apj, 697, 1918 

Marqu{\'e}, C.\ 2004, \apj, 602, 1037 

Newkirk, G., Jr.\ 1961, \apj, 133, 983 

Patsourakos, S., Vourlidas, A., \& Stenborg, G.\ 2013, \apj, 764, 125 

Plowman, J., Kankelborg, C., \& Martens, P.\ 2013, \apj, 771, 2 

Saito, K., 1970, Ann. Tokyo Astron. Obs., Ser 2, 12, 53

Sheeley, N.~R., Jr., Warren, H.~P., \& Wang, Y.-M.\ 2007, \apj, 671, 926 

Shibasaki, K., Alissandrakis, C.~E., \& Pohjolainen, S.\ 2011, Sol. Phys., 273, 309 

Subramanian, P., \& Dere, K.~P.\ 2001, \apj, 561, 372 

Uralov, A.~M., Lesovoi, S.~V., Zandanov, V.~G., \& Grechnev, V.~V.\ 2002, 
Sol. Phys., 208, 69 

Williams, D.~R., T{\"o}r{\"o}k, T., D{\'e}moulin, P., van Driel-Gesztelyi, L., 
\& Kliem, B.\ 2005, \apjl, 628, L163 

Zhang, J., Kundu, M.~R., White, S.~M., Dere, K.~P., \& Newmark, J.~S.\ 2001, \apj, 561, 396 

Zhang, J., Cheng, X., \& Ding, M.-D.\ 2012, Nature Communications, 3,  


\end{document}